\documentclass{desyproc}

\begin{document}
\title{Dark Radiation from a hidden U(1)}

\author{{\slshape Hendrik Vogel$^1$, Javier Redondo$^{1,2}$}\\[1ex]
$^1$Max Planck Institute for Physics, Munich, Germany\\
$^2$Arnold Sommerfeld Center, Ludwig-Maximilians-Universit\"{a}t, Munich, Germany}

\contribID{Vogel\_Hendrik}

\desyproc{DESY-PROC-2014-XX}
\acronym{Patras 2014} 
\doi 

\maketitle

\begin{abstract}
We discuss the impact of a hidden sector consisting of Minicharged
Particles (MCPs) and massless hidden photons on the expansion history of
our Universe. We present parameter scans for the amount of extra relativistic particles
($N_{\text{eff}}$) and the abundance of light nuclei for fermionic MCPs with masses between $\sim$100 keV
and 10 GeV and minicharges in the range $10^{-11}-1$. Current CMB and BBN data significantly constrain the available parameter space of MCPs. The shown results are a valuable indicator for future experimental searches and are presented in a flexible way so that more accurate results on $N_{\rm eff}$ can be easily interpreted. 
\end{abstract}

\section{Minicharged particles}

Minicharged Particles (MCPs) naturally arise in extensions of the Standard Model (SM) with an additional local hidden gauge group $U(1)_{\rm h}$. The hidden photon (HP) associated with this $U(1)_{\rm h}$ does not couple to any SM particle but mixes kinetically with the SM photon
$$
\mathcal{L} = -\frac{\chi}{2}F_{\mu \nu} F'^{\mu \nu},
$$
where $\chi$ is the kinetic mixing parameter, $F_{\mu \nu}$ is the photon field strength tensor and $F'^{\mu \nu}$ is the field strength tensor of the HP. We assume that the $U(1)_{\rm h}$ is unbroken. Hence, the HP is massless. It would be unobservable because the mixing can be removed through a field redefinition. The kinetic mixing becomes unphysical.\\
The HP field has observable consequences in a minimal extension of this model with a (fermionic) field $f$, the hidden fermion. This fermion is only charged under the $U(1)_{\rm h}$ with a coupling strength $g'$. This makes $\chi$ a physical parameter because transforming the kinetic mixing away induces an effective coupling of the hidden fermion to the SM photon with a coupling strength $g'\chi$. Since the hidden fermion couples to the SM photon, it carries an effective electric charge. It is useful to define 
$$
g'\chi=e\epsilon,
$$
where $e$ is the electron charge and $\epsilon$ is the minicharge. Since $\epsilon$ can be small, $f$ is called a minicharged particle. It is fully characterized by $g'$, $\epsilon$ and its arbitrary mass $m_f$.\\

\section{Impact on the CMB and BBN}

\begin{figure}[!ht]
\centerline{\includegraphics[width=0.45\textwidth]{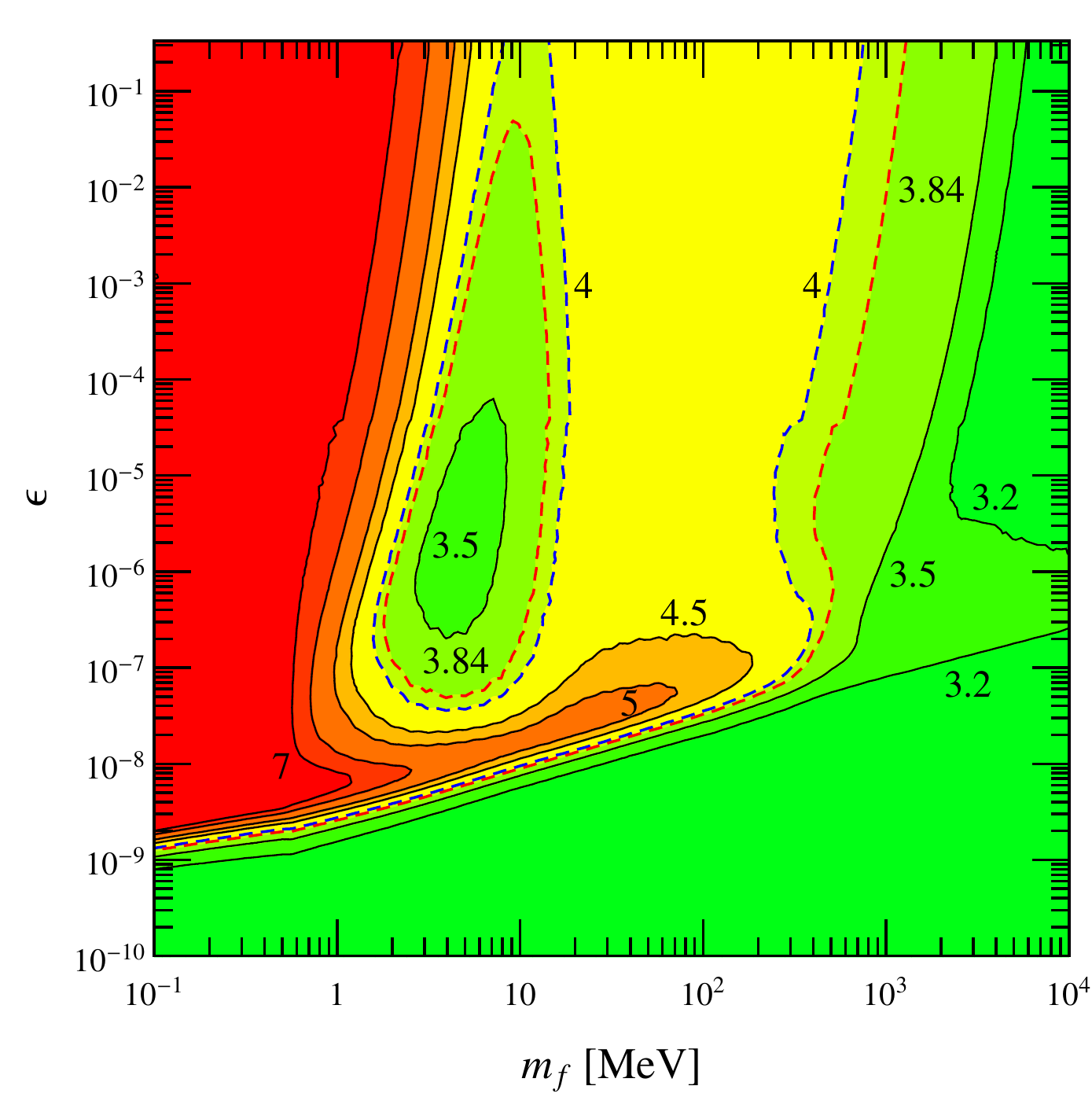}\includegraphics[width=0.45\textwidth]{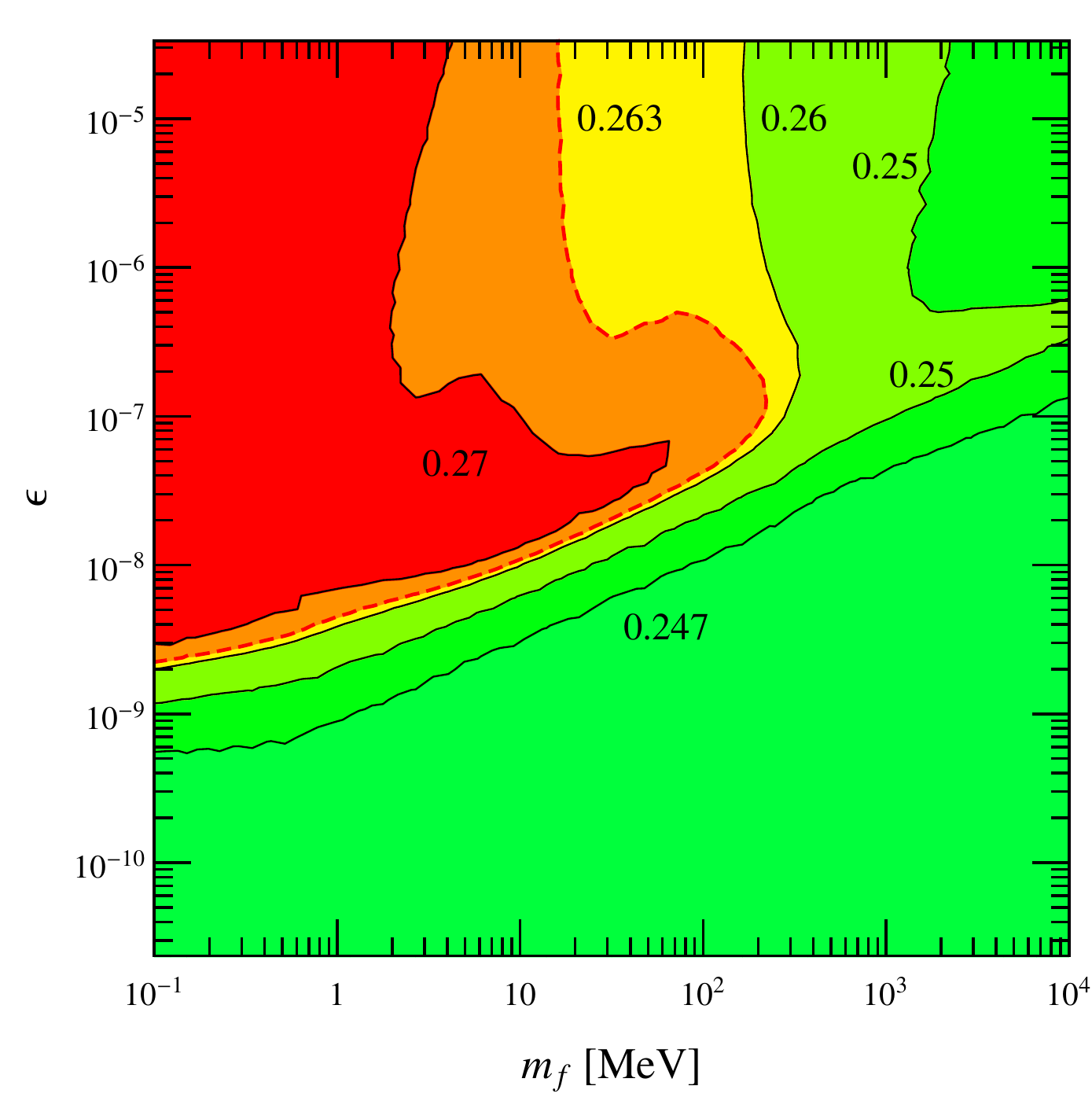}}
\caption{Isocontours of $N_{\rm eff}$ at the CMB epoch (left) and primordial Helium yield $Y_p$ (right) as a function of $m_f$ and $\epsilon$, for $g'=0.1$ . Numbers correspond to the value of the closest contour line. \emph{Left:} Dark green: $N_{\text{eff}}\sim 3$, light green and yellow $N_{\text{eff}}\sim 3.5-4.5$,
Orange and red $N_{\text{eff}}>4.5$. 
The red dashed line shows the 95\% upper exclusion limit $N_{\text{eff}}=3.84$ (Planck+WP+highL+BAO) by Planck~\cite{Ade:2013zuv}. The blue dashed line gives the best fit value $N_{\text{eff}}= 4$ from a combination of BICEP2 and Planck data~\cite{Giusarma:2014zza}. \emph{Right:} Dark green coloring denotes regions far away from the upper limit $Y_p<0.263$~\cite{Mangano:2011ar}. The limit is given by the red dashed contour line. Orange and red regions are excluded on more than a 95\% CL.}\label{Fig:CMB}
\label{sec:figures1}
\end{figure}

Due to the small charge, HPs and MCPs will be produced in the early Universe to some extent. Both of them will then contribute to the energy density of radiation during the formation of the CMB anisotropies. The impact of light degrees of freedom on the CMB is parametrized with the effective neutrino degrees of freedom $N_{\rm eff}$. They are defined as
$$
N_{\rm eff} = \frac{\rho_{\rm DR}}{\rho_{1 \nu*}},
$$
where $\rho_{\rm DR}$ is the energy density of dark radiation (DR), i.e. all relativistic particles that are not strongly coupled to the SM photon at the CMB epoch, $\rho_{1\nu*}$ is the energy density of one neutrino species which instantly decoupled from the electron-photon plasma before BBN. All densities are evaluated at times corresponding to the formation of the CMB anisotropies. $N_{\rm eff}$ scales like $(T_{\rm DR}/T_\gamma)^4$. It is, hence, very sensitive to temperature changes. \\
In the SM the three neutrinos contribute $N_{\rm eff} = 3.046$~\cite{Mangano:2005cc} after correcting for non-instantaneous decoupling. The most precise observations to date come from the PLANCK collaboration. They set an upper limit of $N_{\rm eff}<3.84$ at 95 \% CL on the number of relativistic degrees of freedom at the CMB epoch. This value is obtained by fitting a Cosmological model to the data. The limit is conservative since it does not use the controversial value for the Hubble rate today, $H_0$, obtained from local astrophysical measurements. Local measurements prefer a higher value for $N_{\rm eff}$. Also the disputed result obtained by BICEP2~\cite{Ade:2014xna} suggests a higher value, $N_{\rm eff}= 4$~\cite{Giusarma:2014zza}.\\
In order to compare the predictions of models with minicharged particles to the observations, we have to accurately compute the energy density of dark radiation at the CMB epoch. We numerically solve a set of Boltzmann-like equations that parametrize the energy transport between SM particles and the dark sector (DS) consisting of HPs and MCPs
\begin{eqnarray}
&\dot \rho_{\rm SM} + 3H\left(\rho_{\rm SM}+P_{\rm SM}\right)=-\mathcal{W}\nonumber,\\
&\dot \rho_{\rm DS} + 3H\left(\rho_{\rm DS}+P_{\rm DS}\right)=\mathcal{W}\nonumber,
\end{eqnarray}
where $\rho \ (P)$ is the energy density (pressure) of the DS and the SM particles, $H$ is the Hubble parameter, the dot is a derivative with respect to physical time and $\mathcal{W}$ is a generalized collision term that gives the energy transported from one sector to the other. \\
For the computation we assume that the DS is always in thermal equilibrium with itself. It is fully characterized by a common temperature $T_{\rm DS}$ and $m_f$. Following the spirit of a dark, weakly coupled sector we initially set $T_{\rm DS}=0$. The DS is then produced by the SM particles during the evolution of the Universe. In order to compute this thermalization as accurately as possible we include all SM particles and light mesons and all relevant processes in our computations. It turns out that for $T_{\rm DS}\ll T_\gamma$ DS particles are most efficiently produced by SM particle pair annihilation ($e\bar e\to f\bar f$). When $T_{\rm DS}$  approaches the SM temperature Coulomb-like scattering ($e f\to e f$), regularized by plasma screening, and (for large $g'$) Compton-like scattering ($\gamma f \to \gamma' f$) become the most important processes.\\
We scan over a wide range of parameters $m_f,\epsilon$ for $g\in \{10^{-2},0.1,1\}$. The constraints for $g'=0.1$ are the most conservative. For $g=10^{-2}$ the results are indistinguishable from the results with $g'=0.1$ because in this case the dominant processes are $\propto e\epsilon=g'\chi$. A change in $g'$ can be compensated by a corresponding change in $\chi$ but leaves $\epsilon$ unchanged. The bounds for $g'=1$ are stronger because Compton-like scattering ($\gamma f \to \gamma' f$), which is $\propto g'^4\chi^2 \sim g'^2 \epsilon^2$, becomes dominant for high $g'$. Increasing $g'$ cannot be compensated by a shift in $\chi$ anymore if $\epsilon$ is fixed. For $g'=1$ the DS and SM particles equilibrate faster.\\
Our result for $g=0.1$ is presented in Fig.~\ref{Fig:CMB} (left). The sudden increase for small kinetic mixing $\epsilon \sim 10^{-8}$ corresponds to partial thermalization of the DS with the SM photon. The behavior for large $\epsilon$ can be understood analytically. The DS fully thermalizes and after decoupling its temperature at the CMB epoch is given by entropy conservation. For more details see~\cite{Vogel:2013raa}.\\
Another primordial probe is BBN. The DS changes the formation of nuclei. The most sensitive probe is the yield of primordial helium $Y_p$. By adjusting the BBN code of~\cite{Cadamuro:2011fd}, we computed $Y_p$ using the data of our CMB simulation. The result is shown in Fig.~\ref{Fig:CMB} (right). We use $Y_p<0.263$ at 95\% CL~\cite{Mangano:2011ar} as a conservative upper bound.

\section{Conclusions}

\begin{figure}[!ht]
\centerline{\includegraphics[width=0.6\textwidth]{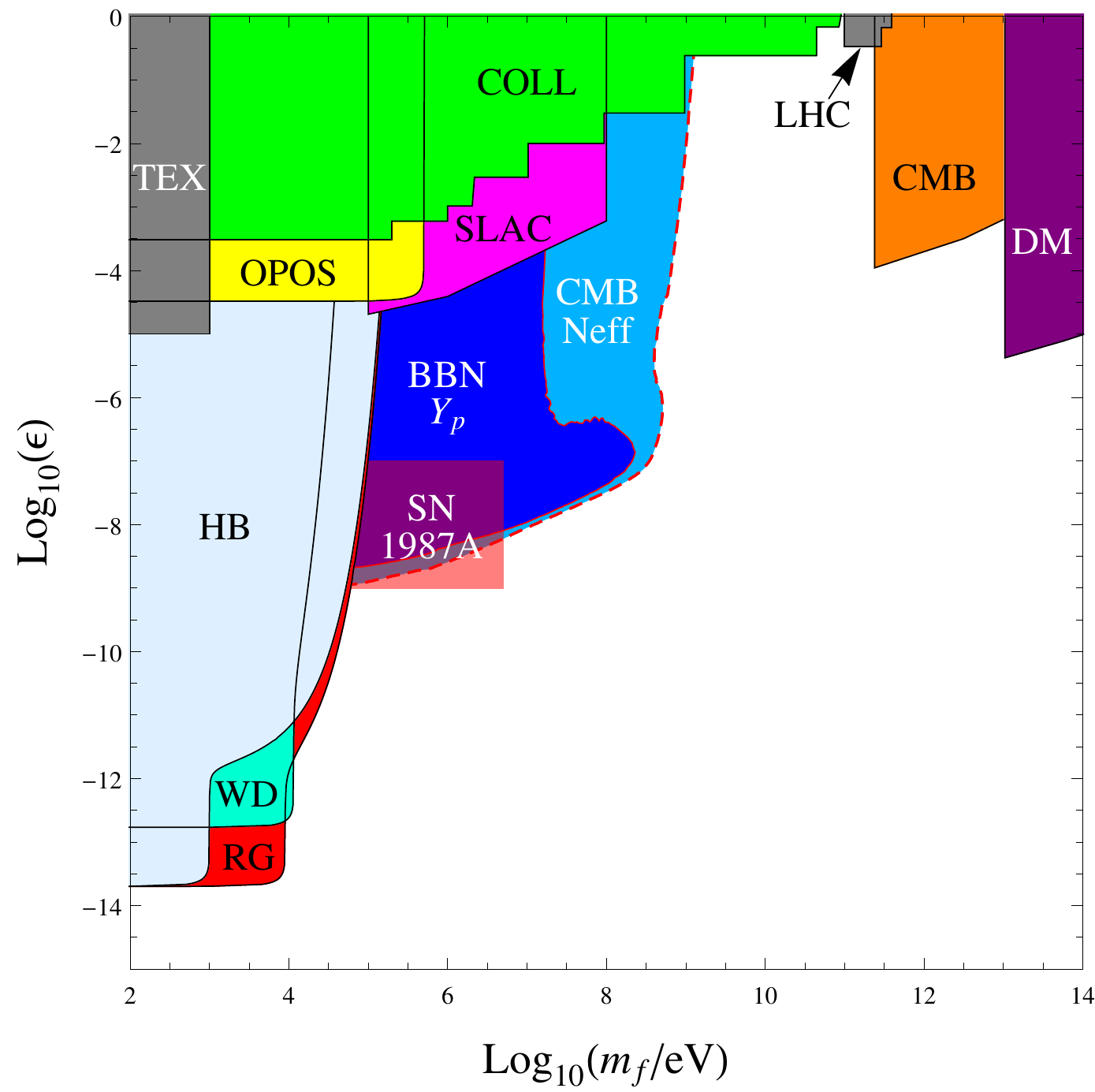}}
\caption{Exclusion plot for MCPs from various experiments and observations. The constraint from the amount of helium $Y_p$ produced during BBN (dark blue) and from light extra degrees of freedom $N_{\text{eff}}$ by Planck (light blue) described here have been obtained in~\cite{Vogel:2013raa}.}\label{Fig:CompleteResult}
\label{sec:figures3}
\end{figure}

Figure~\ref{Fig:CMB} shows that, using PLANCK's limit, the CMB anisotropies allow for light minicharged particles in the range $m_f\sim 5$ MeV for a wide range of minicharges $\epsilon$. This parameter space is, however, disfavored by the abundance of primordial helium. Using these two results we set a lower limit on the hidden fermion mass of $m_f \gtrsim 390$ MeV for $\epsilon > 10^{-7}$. A compilation of this result with earlier approaches using astrophysics, Cosmology or laboratory experiments is shown in Fig.~\ref{Fig:CompleteResult}.\\ If BICEP2 has observed primordial gravitational waves, the limit on MCPs from the CMB anisotropies vanishes and BBN gives the most stringent bounds. Further research is needed to clarify the role of BICEP's observations. Since the question of the existence of additional light degrees of freedom cannot be settled with the current data, we present our results in a flexible way. Future more accurate values of $N_{\rm eff}$ are to be expected. Using the predicted 
accuracy of CMBPol~\cite{Galli:2010it}, PLANCK's current mean value $N_{\rm eff}=3.30$~\cite{Ade:2013zuv} could be detected at $\sim 5\sigma$.

\begin{footnotesize}

\end{footnotesize}

\end{document}